\documentclass[aps,prb,twocolumn,showpacs,superscriptaddress,floatfix,10pt]{revtex4-2}
\usepackage{amsmath,amssymb,amsfonts,stmaryrd,wasysym,graphicx,multirow,color,textcomp,subfigure}
\usepackage{url}
\usepackage[colorlinks=true,citecolor=blue,urlcolor=blue]{hyperref}
\usepackage{romannum}

\normalfont
\def\be{\begin{equation}}
	\def\ee{\end{equation}}
\def\bea{\begin{eqnarray}}
	\def\eea{\end{eqnarray}}

\newcommand{\cyan}[1]{{\color{cyan}{\em {#1}}}}

\newcommand{\vv}[1]{{\boldsymbol #1}}

\begin{document}
	\title{Hinge Magnons from Non-collinear Magnetic Order in Honeycomb Antiferromagnet}
	
	\author{Moon Jip Park}
	\email{moonjippark@kaist.ac.kr}
	\affiliation{Department of Physics, KAIST, Daejeon 34141, Republic of Korea}
	
	\author{SungBin Lee}
	\email{sungbin@kaist.ac.kr}
	\affiliation{Department of Physics, KAIST, Daejeon 34141, Republic of Korea}
	
	\author{Yong Baek Kim}
	\email{ybkim@physics.utoronto.ca}
	\affiliation{Department of Physics, University of Toronto, Toronto, Ontario M5S 1A7, Canada}

	\begin{abstract}

		We propose that non-collinear magnetic order in quantum magnets can harbor a novel higher-order topological magnon phase with non-Hermitian topology and hinge magnon modes. We consider a three-dimensional system of interacting local moments on stacked-layers of honeycomb lattice. It initially favors a collinear magnetic order along an in-plane direction, which turns into a non-collinear order upon applying an external magnetic field perpendicular to the easy axis. We exploit the non-Hermitian nature of the magnon Hamiltonian to show that this field-induced transition corresponds to the transformation from a topological magnon insulator to a higher-order topological magnon state with a one-dimensional hinge mode. As a concrete example, we discuss the recently-discovered monoclinic phase of the thin chromium trihalides, which we propose as the first promising material candidate of the higher-order topological magnon phase. 
	\end{abstract}
	
	\maketitle
	
	\cyan{Introduction} -Topological excitations in quantum magnets have emerged as novel platforms for potential applications in spintronics and quantum information technology\cite{Chumak2015}.~A prominent recent path in pursuit of this direction is the research on topological phases of magnon excitations \cite{PhysRevLett.104.066403,PhysRevX.8.041028,Owerre_2016,doi:10.1063/1.4959815,PhysRevB.100.144401,mook2020interactionstabilized,PhysRevB.87.174427}.~There has been intensive research effort to discover topological magnons in both gapless and gapped phases. Some examples of topological gapless excitations include the topological point and line-nodal magnons \cite{PhysRevB.3.157,PhysRevB.92.144404,PhysRevLett.123.227202,lu2018magnon,PhysRevLett.119.247202,PhysRevX.8.041028,PhysRevB.101.100405}. The candidate materials are CrBr$_3$\cite{PhysRevB.3.157}, Cr$_2$Si$_2$Te$_6$\cite{PhysRevB.92.144404}, three-dimensional Kitaev material, $\beta$-Li$_2$IrO$_3$\cite{PhysRevLett.123.227202}, $\alpha$-RuCl$_3$\cite{lu2018magnon}, and the three-dimensional antiferromagnet Cu$_3$TeO$_6$\cite{PhysRevLett.119.247202}. Gapped magnon spectrum may also carry non-trivial bulk topology, which physically manifests as the boundary magnon modes. The promising candidate materials are the layered transition metal trihalides, CrI$_3$\cite{PhysRevX.8.041028}, and Kagome magnet, YMn$_6$Sn$_6$\cite{PhysRevB.101.100405}. 
	
	The magnonic topological phases have often been understood in analogy with the counterparts of electronic topological systems. However, the topological magnons in quantum magnets, in principle, may involve more complex physical phenomena that the electronic analogy may not be applicable. For example, the topological magnons in non-collinear order are described by intrinsically non-Hermitian Hamiltonian, which may lead to a variety of different topological phases\cite{PhysRevLett.120.146402,PhysRevLett.122.076801,PhysRevLett.124.086801,PhysRevLett.102.065703,PhysRevLett.121.086803,PhysRevB.99.201411}. In particular, such phases may support higher-order topological magnon phases, characterized by gapless ($d-2$)-dimensional boundary excitations in $d$-dimensional bulk.

\begin{figure}[t!]
		\includegraphics[width=0.5\textwidth]{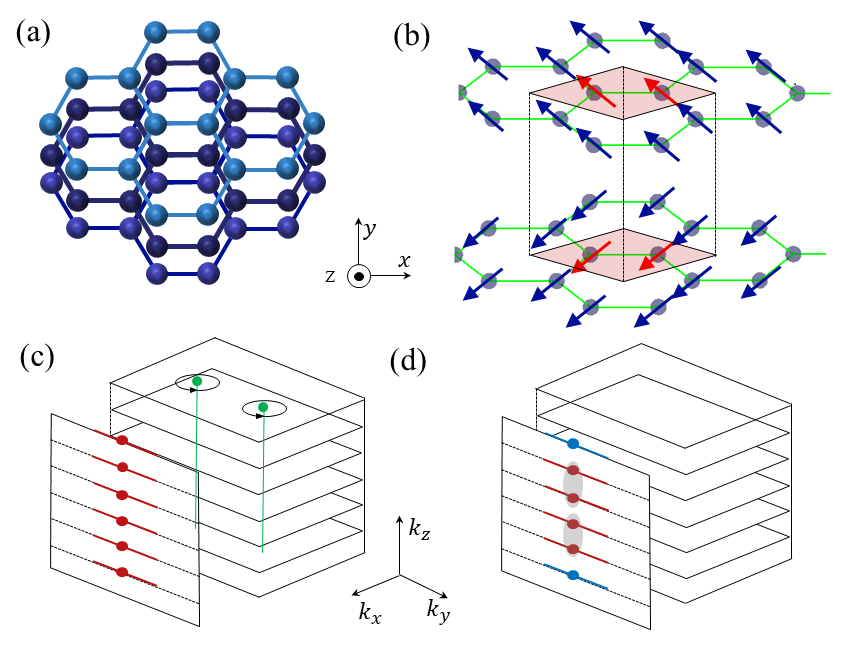}
		\caption{(a) Top view of the atomic configurations in the monoclinic stacked-honeycomb-lattices. (b) Non-collinear magnetic order driven by an external magnetic field. Red spins represent the magnetic unit cell. (c) Locations of magnon edge modes in the Brillouin zone for the monoclinic structure, allowed by the spin-space group. Green points represent the nodal points protected by $\mathbb{Z}_2$-quantized Berry phase with the effective $PT$ symmetry. The projection to the surface perpendicular to the $x$-axis realizes the zigzag edge modes (red lines), which form an effective spin chain along the $z$-direction (red dots). (d) In the non-collinear phase, the DMI dimerizes the edge modes in the effective Kitaev chain along the $z$-direction, leaving only the hinge magnon states (blue lines).
		}
		\label{Fig1}
	\end{figure}

In this work, we present a theoretical study of the magnonic higher-order topological excitations in non-collinear antiferromagnetic order. As a concrete example, we focus on a theoretical model relevant for the recently-discovered monoclinic antiferromagnetic phase of thin chromium trihalides, CrI$_3$\cite{Ubrig_2019,McGuire2015,Song1214,Huang2017,Klein1218,Wang2018,Kim2018} and CrCl$_3$\cite{Klein2019}, where Cr local moments interact with each other in a three-dimensional system of stacked honeycomb lattices. It is shown that the external magnetic field perpendicular to the easy axis drives a phase transition from a collinear order to a non-collinear antiferromagnetic state (Fig.\ref{Fig1} (b)), where the paraunitarity of the magnon wave function introduces the intrinsic non-Hermiticity in the magnon Hamiltonian.  \cite{PhysRevB.100.100405,PhysRevB.100.104423}. In the non-collinear phase, it is shown that the simple analogy with electronic states does not apply. We find that the anomalous non-Hermitian terms in the model gap out the two-dimensional surface mode and generate one-dimensional hinge magnon mode \cite{mook2020chiral}. Finally, we propose the symplectic Wilson loop as a novel bulk topological invariant that correctly captures the hinge magnon modes in the higher-order topological magnon phases\cite{Klein2019}.
	
	\cyan{Spin model} - A single layer of the chromium trihalides forms a simple honeycomb lattice of Cr atoms, which are surrounded by six adjacent non-magnetic halide atoms. The multilayer monoclinic stacking (space group $C2/m$) is obtained by $y$-directional lateral shift to the neighboring layers (See Fig. \ref{Fig1} (a)), which preserves the inversion, $C_{2x}$ rotation and their product, $M_{x}$, mirror symmetry. Motivated by the recent experiments\cite{Ubrig_2019,Klein2019}, we consider a minimal model for the interlayer antiferromagentic order. The spin model consists of general intralayer and interlayer nearest-neighbor couplings that preserve the underlying symmetries of the $C2/m$ group:
	\bea
	H=\sum_{\langle i,j\rangle}J \mathbf{S}_i \cdot \mathbf{S}_j +\sum_{z_i=z_j+1}J_\perp \mathbf{S}_i \cdot \mathbf{S}_j + \mathbf{D}_{\perp,ij}\cdot \mathbf{S}_i \times \mathbf{S}_j,
	\nonumber
	\\
	\label{Eq:spinH}
	\eea
	where $J<0$ and $J_\perp>0$ represent the intralayer ferromagnetic and interlayer antiferromagnetic Heisenberg interactions respectively. As long as the nearest neighbor couplings are concerned, the inversion and $C_{2x}$ symmetries only allow the interlayer Dzyaloshinski-Moriya interaction (DMI), the direction of which lies on the $y-z$ plane $(\mathbf{D}_{\perp}\perp\hat{\mathbf{x}})$, with the opposite signs for the two honeycomb sublattices (A,B). In addition, we assume the presence of a small single-ion anisotropy that stabilizes the antiferromagnetic order along an easy-axis\cite{doi:10.1080/00268976300100491}.
	
	In the AA or rhombohedral (space group $R\bar3$) stacking, the additional $C_{2y}$ and $C_{3z}$ symmetries prohibit the interlayer DMI. The next-nearest neighbor DMI appears as the next leading order coupling\cite{PhysRevX.8.041028}. In such a case, each layer realizes the magnonic analogue of the Haldane model, characterized by the non-trivial Chern number \cite{PhysRevLett.61.2015,Owerre_2016,doi:10.1063/1.4959815}. In this work, we focus on the symmetry-allowed nearest neighbor interlayer couplings of the monoclinic stacking under $C2/m$ group.

	With a given magnetic ground state, the bosonic Bogoliubov Gennes (BdG) Hamiltonian can be derived by using the Holstein-Primakoff (HP) transformation\cite{PhysRev.58.1098}: $S_{+(-)} \approx \sqrt{2S}a^{(\dagger)}$, $S_z \approx S-a^\dagger a$.  The BdG Hamiltonian can be diagonalized via $	H_{\textrm{BdG}}(\mathbf{k})=U_\mathbf{k} D_\mathbf{k}U_\mathbf{k}^\dagger$, where $D_\mathbf{k}$ is the diagonal matrix containing the magnon band energy. In the presence of the $U(1)$-spin rotation symmetry along the easy axis, the Bogoliubov-de Gennes (BdG) Hamiltonian can be decoupled into two sectors as, 
	\bea
	H_{\textrm{BdG}}(\mathbf{k})=H_{s=+1}\oplus H_{s=-1},
	\eea
	where each sector consists of the Hamiltonian of the magnons carrying spin $\pm1$ along the collinear axis respectively. In this case, the wave functions satisfy the unitary condition, $U_\mathbf{k}^\dagger U_\mathbf{k}=I$, as the electron wave function does. However, in general non-collinear orders, the unitary condition is not satisfied. Instead, the bosonic commutation relation demands the paraunitary condition of the wave functions such that $U_\mathbf{k}^\dagger\Sigma_z U_\mathbf{k}=\Sigma_z$, where $\Sigma_z$ is the Pauli matrix acting on the particle-hole space. Obtaining the paraunitary wave function requires the diagonalization of the non-Hermitian Hamiltonian, $\Sigma_{z}H_{\textrm{BdG}} (\mathbf{k})$\cite{COLPA1978327}.

	\cyan{First-order Topology in Collinear order} - When the spin-orbit coupling is negligibly small, the dominant interactions may approximately preserve higher symmetry operations than is strictly required by the magnetic space group. The set of these operations is formally referred as the spin-space group\cite{doi:10.1098/rspa.1966.0211,doi:10.1063/1.1708514}. The spin-space group consists of larger rotational symmetries than that of the magnetic space group\cite{PhysRev.134.A1602}, which independently act on the coordinate rotations and the spin rotations. Therefore, under the spin-space group, we can define the effective space-time inversion symmetry, $PT$, which provides the topological protection of the $\mathbb{Z}_2$-quantized Berry phase along a closed-loop in Brillouin zone(BZ) \cite{PhysRevB.82.115120,Ryu_2010}. 
	
	Without the interlayer interactions, each layer possesses the Dirac nodal points at $K$ and $K'$ points as the monolayer graphene does (See Fig. \ref{Fig1}(c)). The topological protection of the nodal point can be formally cast by $\mathbb{Z}_2$-quantized Berry phase $\pi$ along a loop encircling the nodal points (Black circles in Fig. \ref{Fig1}(c)). The physical manifestation of $\pi$-Berry phase is the zigzag edge modes at the two-dimensional surfaces of the three-dimensional bulk \cite{Li2017,PhysRevB.93.121113}. In the presence of non-negligible spin-orbit coupling, DMI may opens a band gap, leading to the magnonic topological insulator with a non-trivial Chern number \cite{PhysRevX.8.041028}.
	
	\cyan{Second-order Topology in Non-collinear order} - We now consider the external magnetic field along the direction perpendicular to the collinear antiferromagnetic order. That is, we add the term, $H_{\textrm{ext}}=-\mathbf{h}_{\textrm{ext}} \cdot\sum_i \mathbf{S}_i$, where $\mathbf{h}_{\textrm{ext}}$ represents the external magnetic field. The magnetic field generates the non-collinear order with the spin canting (See Fig. \ref{Fig1} (b)). In this case, the conventional Wilson line and the Berry phase are not well-defined anymore, since the wave function does not satisfy the unitary condition\cite{COLPA1978327}. Instead, we show that the non-collinearity induced by the spin canting generates the higher-order topological hinge magnon states (See Fig. \ref{Fig1}(d)).

\begin{figure}[t]
		\includegraphics[width=0.5\textwidth]{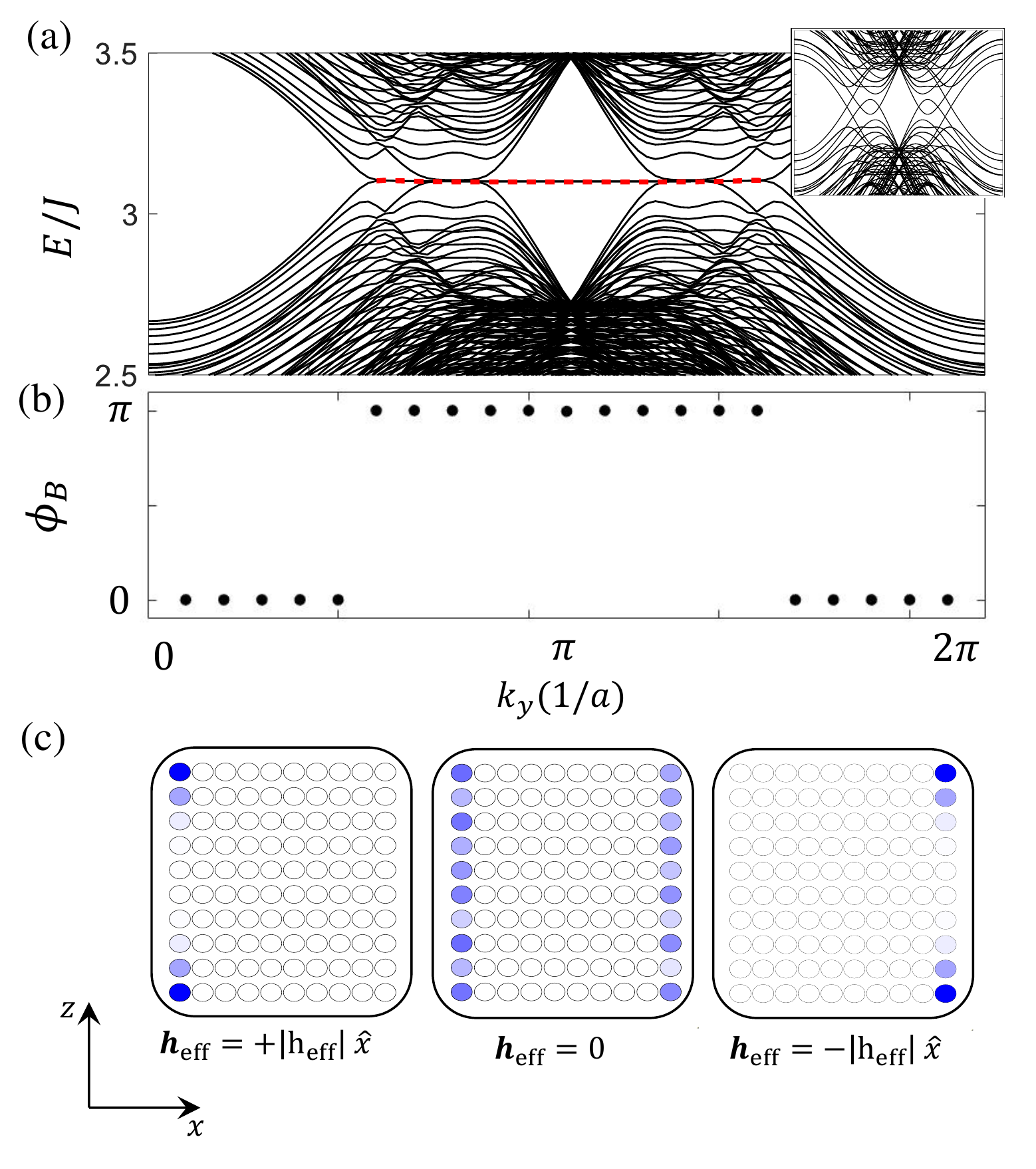}
		\caption{ (a) The magnon band structure as a function of $k_y$ with open boundary condition along $x-z$ direction. In the middle of the bands, the hinge magnon states emerge  (red lines). Inset: the same band structure with the periodic boundary condition along the $z$-direction. Here, the DM vector is pointing in the $z$-direction. (b) The non-Abelian Berry phase as a function of $k_y$. We find the $\pi$-quantized Berry phase at $k_y=\pi$, where the hinge modes reside. (c) The wave function character of the hinge mode at $k_y=\pi$. We find that the wave functions are localized at the corner of the $x-z$ plane. 
 		}
		\label{Fig2}
	\end{figure}

	As an example, we first consider the specific case where the $x$-directional magnetic field is applied to the $y$-directional antiferromagnetic order. Fig. \ref{Fig2}(a) shows the non-collinear magnon bands. We find that the zigzag edge modes at the surface are immediately gapped out as the magnetic field is applied (Inset of Fig. \ref{Fig2}(a)). In contrast, a pair of in-gap states emerge within the momentum window $K<k_y <K'$ (Red dashed line in Fig. \ref{Fig2}(a)), as we take the additional open boundary condition along the $z$ direction. Unlike the surface states, these in-gap states are localized at the corner of the $x-z$ plane, evidently showing the nature of the hinge modes (Fig. \ref{Fig2} (c)). In addition, unlike the conventional electronic higher-order topological insulators, the localization of the hinge modes occurs only at the two corners of one side surface. The surface possessing the hinge mode switches as the direction of the magnetic field is inverted.
	
\begin{figure}[t!]
		\includegraphics[width=0.5\textwidth]{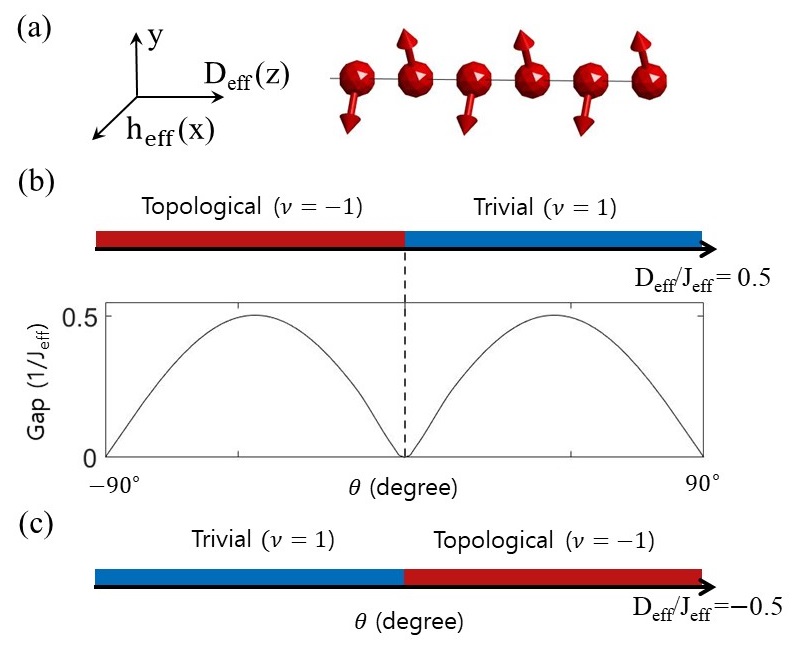}
\caption{(a) Schematic picture of the one-dimensional spin chain along the $\hat{z}$-direction. The external magnetic field is applied along the $\hat{x}$-direction, while the DM vector is pointing in the $\hat{z}$-direction. (b) The band gap and topological phase diagram as a function of the external magnetic field and the DMI. The topological phase (red strip) is characterized by the topological number $\nu=-1$ ($\pi$ Berry phase). The physical manifestation is the zero-dimensional bound mode at the end of the chain. (c) The same phase diagram as in (b) but the direction of the DMI is reversed.}
		\label{Fig3}
	\end{figure}

	\cyan{Effective surface model}- A heuristic way of understanding the emergence of the hinge state is to consider the effective model describing the zigzag edge modes at the surface (Fig. \ref{Fig3}(a)). At a fixed $k_y$, we can consider the zigzag edges as the effective one-dimensional antiferromagnetic chain ($J_\textrm{eff}>0$) along the $\hat{z}$-direction, 
	\bea
	H_\textrm{eff}=\sum_{\langle i,j \rangle}J_\textrm{eff} \mathbf{S}_i \cdot \mathbf{S}_j + (D_\textrm{eff}\hat{z})\cdot \mathbf{S}_i \times \mathbf{S}_j-(h_\textrm{eff}\hat{x})\cdot\sum_i \mathbf{S}_i,
	\nonumber
	\\
	\label{Eq:1D}
	\eea
	where the subscript indices $i$ indicates the $i$-th layer. The non-collinear order in the presence of the spin canting can be represented with the cant angle $\theta$ as, $\vec{S}_i=|S|(\sin\theta,(-1)^i \cos\theta,0)$. Due to the non-collinearity, the HP transformations of the Heisenberg exchange interaction and the DMI now contain both particle-particle and particle-hole channels as,
	\bea
	\nonumber
	&J_\textrm{eff}& \enspace : \enspace
	\mathbf{S}_{i+1} \cdot \mathbf{S}_{i}
	=
	\\
	&\frac{|S|}{2}\times&
	\big[
	\sin^2 \theta a^\dagger_{i+1} a_{i}-\cos^2 \theta a^\dagger_{i+1} a^\dagger_{i}  -n_{i+1}-n_i]+\textrm{h.c.}
	\nonumber
	\eea 
	\bea
	\label{Eq:HP}
	&D_\textrm{eff}& \enspace : \enspace
	\hat{z}\cdot \mathbf{S}_{i+1} \times \mathbf{S}_i=
	\\
	&\frac{|S|}{2}&\times
	\big[
	\frac{(-1)^{i}}{2} \sin 2\theta(a^\dagger_{i+1} a_{i}+a^\dagger_{i+1} a^\dagger_{i}
	-n_{i+1}-n_i) \big]+\textrm{h.c.}
	\nonumber
	\eea
	where $n_i$ is the HP boson number operator. Here, we have only collected the quadratic terms of the HP bosons within the linear spin-wave theory. The resulting HP Hamiltonian consists of the normal hopping and $p$-wave pairing terms as well as a chemical potential, and resembles the Kitaev chain model\cite{Kitaev_2001} that possesses the Majorana fermions at the boundary. 
	
	We now formally show that the above one-dimensional spin chain is characterized by the non-trivial topology. By collecting the interactions in Eq. \eqref{Eq:HP}, we derive the full tight-binding model of the HP bosons as,
	
	\bea
	\label{Eqs:1Dkitaev}
	H_{\textrm{HP}} &=& \frac{|S|}{2} \sum_{i} 
	t_{\textrm{eff},i}[ a^\dagger_{i+1}a_{i}+a_{i+1}a^\dagger_{i}]
\\	&+&\Delta_{\textrm{eff},i} [a^\dagger_{i+1}a^\dagger_{i}+ a_{i+1}a_{i}] - \mu_{\textrm{eff}} [a^\dagger_i a_i+a_i a^\dagger_i]+\textrm{h.c.},
\nonumber
	\eea
	where the explicit form of the coupling parameters are given by, 
	\bea
	t_{\textrm{eff},i}&=& \frac{J_{\textrm{eff}}}{2}(1-\cos 2\theta) +(-1)^i \frac{D_{\textrm{eff}}}{2} \sin 2 \theta,
	\\
	\Delta_{\textrm{eff},i}&=&-\frac{J_{\textrm{eff}}}{2}(1+\cos2\theta)+(-1)^{i}\frac{D_{\textrm{eff}}}{2} \sin 2\theta,
	\nonumber
	\\
	\mu_{\textrm{eff}}&=&-2J_{\textrm{eff}}\cos2\theta -\frac{h_{\textrm{eff}}}{|S|}\sin\theta.
	\nonumber
	\eea
	The above Hamiltonian can be analytically diagonalized (See Supplementary Material for the detailed analytical calculations). Indeed, we find that it supports two topologically distinct gapped phases: the non-trivial phase with the zero-dimensional boundary mode and the trivial gapped phase. The topological phase transition between the two gapped phases occurs when $\theta=0$, regardless of the specific parameters in the HP Hamiltonian (Fig. \ref{Fig3} (b)). This behavior explains the emergence of the hinge mode even with an arbitrary small strength of the magnetic field. By further increasing the strength of the magnetic field, we find that the magnon band gap closes at the cant angle $\theta=\pm 90^\circ$, where
	the system regains the collinearity along the direction of the magnetic field. 
	
	Furthermore, the non-trivial phase is determined by the relative orientation between the DM vector and the direction of the magnetic field (Fig. \ref{Fig3}(c)).  In the full three-dimensional model, the zigzag edges of the two side surfaces consist of different sublattices, in which the direction of the DMI is reversed from each other. Therefore, at a given direction of the magnetic field, one of the two side surfaces becomes topologically non-trivial, while the other surface is topologically trivial. As a result, only two hinge modes occur at one of the side surfaces, which explains the localization pattern of the hinge mode in the three-dimensional model (Fig. \ref{Fig2}(c)). Finally, we note that our results hold regardless of the specific direction of the collinear magnetic order as far as the external magnetic field is applied in the perpendicular direction. As we show in the next section, the hinge modes are robust as long as the magnetic order preserves $C_{2x}$ symmetry.

	\cyan{Symplectic Willson loop} - To rigorously describe the topology of the paraunitary wave functions, we introduce the topological invariant. The particle-hole symmetry of the BdG Hamiltonian allows the decomposition of the wave functions into the positive and negative energy sectors as $U_\mathbf{k}=(V_\mathbf{k}, \Sigma_x V_{-\mathbf{k}})$ where $V_\mathbf{k}$ is $N\times \frac{N}{2}$-dimensional matrix, containing the eigenvectors of the positive energy sector \cite{PhysRevB.99.041110,Kondo2020,lu2018magnon}. The wave functions of the positive energy sectors satisfy the following normalization condition,
	$
	 V_\mathbf{k}^\dagger \Sigma_z V_\mathbf{k}=I.
	$
	Using this property, we can define the \textit{sympletic Wilson line} as,
	\bea
	\mathcal{U}_s(\vv{k}_1\rightarrow \vv{k}_2) \equiv 
	\Sigma_z
	\hat{P}(\vv{k}_1)[\prod_\vv{k}\Sigma_z \hat{P}(\vv{k})] \Sigma_z \hat{P}(\vv{k}_2),
	\label{Eq:wilson}	
	\eea
	where the momentum vectors, $\mathbf{k}$, form a path from $\vv{k}_1$ to $\vv{k}_2$. $[\hat{P}(\vv{k})]_{ij}=\sum_{n\in \textrm{occupied}} [V_\mathbf{k}]_{i,n} [V_\mathbf{k}^\dagger]_{n,j}$ is the projection operator to the occupied positive energy subspace at the momentum $\vv{k}$. In contrast to the conventional Wilson line, the additional operator, $\Sigma_z$, is inserted between the projection operator. It is important to note that Eq. \eqref{Eq:wilson} along a closed loop recovers the unitarity due to the normalization condition. Utilizing the unitarity condition, we can formally define the non-Abelian Berry phase of the paraunitary wave functions as,
	\bea
	e^{i\Phi_\textrm{B}(\vv{k}_1\rightarrow \vv{k}_2)}=
	\det[\mathcal{U}_s(\vv{k}_1\rightarrow \vv{k}_2)\Sigma_z].
	\eea
	
	In general, the Berry phase can be any arbitrary value between $0$ and $2\pi$ ($\Phi_\textrm{B}\in [0,2\pi)$). However, in the presence of $C_{2x}$ rotation symmetry, the Berry phase becomes quantized. To show this, we consider the one-dimensional Hamiltonian shown in the previous section. We now introduce the following topological invariant, $\nu$. Using $C_{2x}$ symmetry, we can decompose the Wilson loop into the two Wilson lines related by $C_{2x}$ symmetry as,
	\bea
	\nu\equiv e^{i\Phi_B}
	&=& \det [ \mathcal{U}_{s,(0\rightarrow\pi)}  \mathcal{U}_{s,(\pi\rightarrow2\pi)} \Sigma_z]
	\\
	\nonumber
	&=& \det [  \mathcal{U}_{s,(0\rightarrow\pi)} \hat{C}_{2x}^{-1} (\hat{C}_{2x} \mathcal{U}_{s,(\pi\rightarrow 2\pi)} \hat{C}_{2x}^{-1}) \hat{C}_{2x} \Sigma_z
	]
		\\
	\nonumber
	&=& \det [  \mathcal{U}_{s,(0\rightarrow\pi)} \hat{C}_{2x}^{-1} \mathcal{U}_{s,(\pi\rightarrow 0)}  \hat{C}_{2x} \Sigma_z
	]
	\\
	\nonumber 
&=& 	\prod_{n\in\textrm{occupied}} \zeta_{n}(0)\zeta_{n}(\pi),
	\eea
where $\zeta_n(k)$ is the eigenvalue of $C_{2x}$ on $n$-th band at the momentum $k$. Since $\zeta_n=\pm1 $ ($C_{2x}^2=1$), the Berry phase can only be $0$ or $\pi$, which gives rise to $\mathbb{Z}_{2}$-topological protection of the hinge mode. (See Supplementary Material for the detailed mathematical proof).
 
We now numerically calculate the Wilson loop via integration along the $k_z$-direction in the three dimensional model with the open boundary condition in the $x$-direction. Fig. \ref{Fig2} (b) shows the computed Berry phase as a function of $k_y$. We find the quantized $\pi$-Berry phase at $C_{2x}$ invariant momentum, $k_y=\pi$, where the hinge modes appear ($K<k_y <K'$). The $\pi$-quantized Berry phase signals the boundary modes in the $x-z$ plane\cite{RevModPhys.66.899,Benalcazar61}, which physically manifests as the hinge mode.  
	
	\cyan{Discussions} - In conclusion, we have proposed a theoretical model for the higher-order topological hinge magnon excitations in the three-dimensional system of layered honeycomb antiferromagnets. It is shown that the magnetic-field-driven phase transition from a collinear magnetic order to a non-collinear magnetic order, corresponds to the topological phase transition from a magnon topological insulator to a high-order topological magnon state with hinge magnon modes. We show that the non-Hermitian nature of the magnon Hamiltonian and the paraunitarity of the magnon wavefunction play crucial roles here. While the conventional Wilson loop approach fails to capture the topology of the system, we demonstrate that the symplectic Wilson loop method offers an alternative topological invariant. Our results suggest that the external magnetic field can be used as a control knob for the topological phase transition.
	
	We propose the multilayer chromium trihalides as a promising material candidate of the higher-order topological magnon phase. Recent Raman spectroscopy measurement has identified the presence of the monoclinic phase ($C2/m$) in the thin multilayers of CrI$_3$ \cite{Ubrig_2019} and CrCl$_3$\cite{Klein2019}. The interlayer antiferromagnetism has been further confirmed through magneto-optical Kerr effect measurement\cite{Song1214,Huang2017} and tunneling magnetoresistance\cite{Klein1218,Wang2018,Kim2018}, which is consistent with the magnetic ordering pattern in our theoretical model. We may also expect a sizable strength of DMI as in the bulk sample. We have shown that the interlayer stacking order plays a crucial role in the form of the DM interaction and the higher-order topology. Experimentally, the monoclinic phases are stable in the thin film, while the bulk sample undergoes the structural phase transition to the rhombohedral structure ($R\bar3$)\cite{McGuire2015}. Therefore, investigations of magnetic properties by varying the layer thickness would reveal an interesting interplay between the stacking order and topology. In addition, we also suggest that an external elecric field could be used to control the interlayer DMI\cite{PhysRevB.97.054416,doi:10.1021/acs.nanolett.8b01502,doi:10.1063/1.5050447}. In particular, the electric field along the $x$-direction would generate the interlayer DMI by breaking $C_{2y}$ symmetry even in the rhombohedral structure($R\bar3$). 
	
	Finally, it is worthwhile to mention that the family of A$_2$TMO$_3$ compounds (A=Na,Li, TM=transition metal) also contains many monoclinic honeycomb antiferromagnets \cite{Takagi2019,PhysRevLett.108.127204,Lee_2012,STROBEL198890,TAKAHASHI20081518}. A zigzag in-plane order in addition to the interlayer antiferromagnetism has been observed in these materials. In such a case, the additional zone folding occurs along the in-plane directions. The effect of the additional in-plane order on the topology would be an interesting subject of future study.

	\acknowledgments
	
	The numerical calculations of the linear spin-wave theory are performed using Spinw software program \cite{Toth_2015}. M.J.P. and S.L. are supported by the National Research Foundation Grant (NRF-2020R1F1A1073870, NRF- 2020R1A4A3079707). Y.B.K. is supported by the NSERC of Canada and the Center for Quantum Materials at the University of Toronto.
	
	\bibliography{reference}
	
	\clearpage
	\pagebreak
	
	\renewcommand{\thesection}{\arabic{section}}
	\setcounter{section}{0}
	\renewcommand{\thefigure}{S\arabic{figure}}
	\setcounter{figure}{0}
	\renewcommand{\theequation}{S\arabic{equation}}
	\setcounter{equation}{0}
	
\begin{widetext}
\section{Supplementary Material}

\subsection{Topological bound states in the effective one-dimensional spin chain}

In this section, we present the detailed calculation of the effective one-dimensional spin chain in Eq. \eqref{Eq:1D}. For the clarity, we start our discussion by writing Eq. \eqref{Eq:1D} again here. 
\bea
H_\textrm{eff}=\sum_{\langle i,j \rangle}J_\textrm{eff} \mathbf{S}_i \cdot \mathbf{S}_j + (D_\textrm{eff}\hat{z})\cdot \mathbf{S}_i \times \mathbf{S}_j-(h_\textrm{eff}\hat{x})\cdot\sum_i \mathbf{S}_i,
\label{Eqs:1D}
\eea
where $J_{\textrm{eff}}>0$, $D_{\textrm{eff}}$, and $h_\textrm{eff}$ represent the effective Heisenberg interaction, DMI, and the external exchange field interaction respectively. If only $J_{\textrm{eff}}$ exists in Eq. \eqref{Eqs:1D}, the spins form the antiferromagnetic order. In this case, we assume that the collinear antiferromagnetic order points out the $\hat{y}$-direction, $\vec{S}=|S|(0,(-1)^i,0)$, where the application of the external field cants the spin to the $\hat{x}$-direction. The non-collinear order on the $x-y$ plane can be represented as, 
\bea
\vec{S}_{\textrm{global},i}=|S|(\sin\theta,(-1)^i \cos\theta,0),
\label{Eqs:Snon}
\eea
where we use the subscript '$\textrm{global}$' to denote that the spin vector is written in the global Cartesian coordinate basis.
To perform the Holstein-Primakoff transformation, we express the spins in the global Cartesian coordinate basis in terms of the local coordinate basis as,
\bea
\label{Eqs:localS}
\begin{pmatrix}
	S_{\textrm{global},x}\\
	S_{\textrm{global},y}\\
	S_{\textrm{global},z}\\
\end{pmatrix}_{\textrm{i-th site}}=
\begin{pmatrix}
	-(-1)^i\cos\theta & 0 &  \sin \theta \\
	 \sin\theta  & 0 &  (-1)^i\cos\theta \\
	0 & 1 & 0\\
\end{pmatrix}
\begin{pmatrix}
S_{x}\\
S_{y}\\
S_{z}\\
\end{pmatrix}_{\textrm{i-th site}}
\eea
where $\vec{S}$ on the right-side of the equation is represented in the local coordinate basis. Accordingly, each interaction term in Eq. \eqref{Eqs:1D} can be rewritten as,

\bea
J_\textrm{eff} \quad :\quad 
\mathbf{S}_{i+1} \cdot \mathbf{S}_{i}|_{\textrm{global}}
=-\cos 2\theta  ({S}_{x,i+1}S_{x,i}+{S}_{z,i+1}S_{z,i})+(-1)^i\sin 2\theta ( S_{x,i+1} S_{z,i}-S_{z,i+1} S_{x,i})+S_{y,i+1}S_{y,i},
\nonumber
\eea
\bea
D_\textrm{eff} \quad :\quad 
\hat{z}\cdot \mathbf{S}_{i+1} \times \mathbf{S}_i|_{\textrm{global}}
=\cos 2\theta(S_{x,i+1}S_{z,i}-S_{z,i+1}S_{x,i} )+(-1)^i \sin 2\theta(S_{x,i+1}S_{x,i}+S_{z,i+1}S_{z,i}),
\eea
\bea
h_\textrm{eff} \quad :\quad \hat{x}\cdot\mathbf{S}_i|_{\textrm{global}}=-(-1)^i\cos \theta S_{x,i} +\sin \theta S_{z,i}.
\nonumber
\eea
After we express the spin Hamiltonian in the local coordinate basis, we perform the standard HP transformation with the following convention :
$ S_{+} \approx \sqrt{2S}a, \quad S_{-} \approx \sqrt{2S}a^\dagger, \quad S_z \approx S-a^\dagger a $. Each interaction term is rewritten in terms of the HP bosons as,

\bea
J_\textrm{eff} \quad :\quad 
\mathbf{S}_{i+1} \cdot \mathbf{S}_{i}|_{\textrm{global}}
=\frac{S}{2}\times \big[
-\cos 2\theta [ (a+a^\dagger)_{i+1} (a+a^\dagger)_{i}-2(n_{i+1}+n_i)]- (a-a^\dagger)_{i+1} (a-a^\dagger)_{i} \big],
\eea 
\bea
D_\textrm{eff} \quad  : \quad 
\hat{z}\cdot \mathbf{S}_{i+1} \times \mathbf{S}_i|_{\textrm{global}}=
\frac{S}{2}\times
\big[
(-1)^{i} \sin 2\theta[(a+a^\dagger)_{i+1} (a+a^\dagger)_{i} -2(n_{i+1}+n_i)] \big],
\nonumber
\eea
\bea
h_\textrm{eff} \quad : \quad 
\hat{x}\cdot\mathbf{S}_i|_{\textrm{global}}
=-sin \theta n_i.
\nonumber
\eea

Here, we only collect the quadratic terms of the HP bosons within the linear-spin wave theory. Finally, we derive the one-dimensional tight-binding model of the HP bosons as,

\bea
H_{\textrm{HP}}= \frac{S}{2} \sum_{i} 
t_{\textrm{eff}}[ a^\dagger_{i+1}a_{i}+a_{i+1}a^\dagger_{i}]
+\Delta_{\textrm{eff}} [a^\dagger_{i+1}a^\dagger_{i}+ a_{i+1}a_{i}] - \mu_{\textrm{eff}} [a^\dagger_i a_i+a_i a^\dagger_i]+\textrm{h.c.},
\label{Eqs:1Dkitaev}
\eea
where the effective coupling parameters are explicitly written as,
\bea
t_{\textrm{eff},i}= \frac{J_{\textrm{eff}}}{2}(1-\cos 2\theta) +(-1)^{i} \frac{D_{\textrm{eff}}}{2} \sin 2 \theta,
\\
\Delta_{\textrm{eff},i}=\frac{J_{\textrm{eff}}}{2}(-1-\cos2\theta)+(-1)^{i}\frac{D_{\textrm{eff}}}{2} \sin 2\theta,
\nonumber
\\
\mu_{\textrm{eff}}=-2J_{\textrm{eff}}\cos2\theta-\frac{h_{\textrm{eff}}}{S}\sin\theta.
\nonumber
\eea

We recognize that the tight-binding model of Eq. \eqref{Eqs:1Dkitaev} resembles the one-dimensional Kitaev chain with $p$-wave superconductivity. We can analytically diagonalize Eq. \eqref{Eqs:1Dkitaev}. The corresponding energies are given as,
\bea
E_{\pm,\pm}=\pm \sqrt{\mu_{\textrm{eff}}^2+t_0^2+t_1^2-\Delta_0^2-\Delta_1^2 \pm 2\sqrt{
\mu_{\textrm{eff}}^2(t_0^2+t_1^2)-(t_0\Delta_1+t_1\Delta_0)^2}},
\eea
where $t_{0}(k)=t_{\textrm{eff},2n}+\cos(k) t_{\textrm{eff},2n+1}$, $t_{1}(k)=\sin(k) t_{\textrm{eff},2n+1}$, $\Delta_{0}(k)=\Delta_{\textrm{eff},2n}+\cos(k) \Delta_{\textrm{eff},2n+1}$, $\Delta_{1}(k)=\sin(k) \Delta_{\textrm{eff},2n+1}$. Among the total four bands, the zero energy band gap between the particle-hole symmetric pair is shown to be always topologically trivial  \cite{lu2018magnon}. In contrast, the energy gap between the bands with the positive energies supports the two topologically distinct gapped phases. The gap between the two bands closes if the following condition is satisfied:
\bea
\mu_{\textrm{eff}}^2(t_0^2+t_1^2)-(t_0\Delta_1+t_1\Delta_0)^2=0
\label{Eqs:cond}
\eea
We find that Eq. \eqref{Eqs:cond} is satisfied when the cant angle reaches 
\bea
\theta=0,\pm\pi/2,
\eea
regardless of the specific parameters of the Hamiltonian.

\subsection{Quantization of Berry phase}

In this section, we show that the $C_{2x}$ symmetry gives rise to the quantization of the Berry phase. The non-collinear spin order in Eq. \eqref{Eqs:Snon} satisfy the $C_{2x}$ symmetry acting on the center of the bond center of the one-dimensional spin chain as,
\bea
C_{2x} \quad : \quad S_{x,i}|_{\textrm{global}} &\rightarrow& S_{x,-i+1}|_{\textrm{global}},
\\
\nonumber
 S_{y,i}|_{\textrm{global}} &\rightarrow& -S_{y,-i+1}|_{\textrm{global}},
\\
\nonumber
 S_{z,i}|_{\textrm{global}} &\rightarrow& -S_{z,-i+1}|_{\textrm{global}}.
\eea
In the local coordinate basis, the spins transform as,
\bea
C_{2x} \quad : \quad S_{x,i} &\rightarrow& -S_{x,-i+1},
\\
\nonumber
S_{y,i} &\rightarrow& -S_{y,-i+1},
\\
\nonumber
S_{z,i} &\rightarrow& S_{z,-i+1}.
\eea
From the above relation, we can write $\hat{C}_{2x}$ operator acting on the HP bosons as,
\bea
\hat{C}_{2x} a_{\textrm{even},k} \hat{C}_{2x}^{-1}= -a_{\textrm{odd},-k}
\\
\nonumber
\hat{C}_{2x} a_{\textrm{odd},k} \hat{C}_{2x}^{-1}= -a_{\textrm{even},-k}
\eea
where $a_{\textrm{even(odd)},k}$ is the HP bosons in the even(odd) sites at the momentum $k$ respectively. It is important to note that $[\Sigma_z,\hat{C}_{2z}]=0$ holds. Using this property, we can define the sewing matrix of the $\hat{C}_{2x}$ operator, which relates the magnon eigenstates with its $C_{2x}$ partner. The definition of the sewing matrix can be written as,
\bea
\mathcal{B}_{mn}(k)=\langle u_{m}(-k) | \hat{C}_{2x}\Sigma_z | u_{n}(k) \rangle,
\eea
where $| u_{n}(k) \rangle$ is $n$-th magnon eigenstates at the momentum $k$. At the $C_{2x}$ invariant momenta, we can write as,
\bea
\det[\mathcal{B}(k=0,\pi)]=\prod_{n\in\textrm{occupied}} \zeta_{n}(k=0,\pi),
\eea
where $\zeta_n(k)$ is the eigenvalue of $\hat{C}_{2x}$ of the $n$-th band at the $C_{2x}$ invariant momentum $k$.
We now consider the Wilson loop evaluated along the one-dimensional BZ as,
\bea
(U_s)_{mn} &\equiv&  \sum_{a,b,..\in \textrm{occupied}}\langle u_{m}(0) |\Sigma_z |u_{a}(\mathbf{k}_1)\rangle
\langle u_{a}(\mathbf{k}_1) |\Sigma_z |u_{b}(\mathbf{k}_2)\rangle
\langle u_{b}(\mathbf{k}_2) |\Sigma_z... |u_{n}(2\pi)\rangle
\\
\nonumber
&=& 
\langle u_{m}(0) |\mathcal{U}_{(0\rightarrow 2\pi)}\Sigma_z |u_{n}(2\pi)\rangle,
\eea
where $\mathbf{k_1},...,\mathbf{k_2}$ form a path from $k=0$ to $2\pi$. Here, we ignore the subscript $s$ for the simplicity. Finally, using the property of $\hat{C}_{2x}$ operator, the Berry phase can be simplified as,
\bea
e^{i\Phi_B}&=&\det [\langle u_{m}(0) |\mathcal{U}_{(0\rightarrow2\pi)} |\tilde{u}_{n}(2\pi)\rangle]
\\
\nonumber
&=& \det [\langle u_{m}(0) |  \mathcal{U}_{(0\rightarrow\pi)}  \mathcal{U}_{(\pi\rightarrow2\pi)} |\tilde{u}_{n}(2\pi)\rangle]
\\
\nonumber
&=& \det [\langle u_{m}(0) |  \mathcal{U}_{(0\rightarrow\pi)} \hat{C}_{2x}^{-1} (\hat{C}_{2x} \mathcal{U}_{(\pi\rightarrow 2\pi)} \hat{C}_{2x}^{-1}) \hat{C}_{2x} 
 |\tilde{u}_{n}(2\pi)\rangle]
\\
\nonumber
&=& \det [\langle u_{m}(0) |  \mathcal{U}_{(0\rightarrow \pi)}  \hat{C}_{2x}^{-1}  \mathcal{U}_{(\pi\rightarrow 0)}  \hat{C}_{2x} 
|\tilde{u}_{n}(2\pi)\rangle]
\\
\nonumber
&=& \sum_{a,b,c\in\textrm{occupied}} \det [\langle u_{m}(0) |  \mathcal{U}_{(0\rightarrow \pi)} |\tilde{u}_{a}(\pi)\rangle
\langle u_{a}(\pi) |   \hat{C}_{2x}^{-1}  |\tilde{u}_{b}(\pi)\rangle \langle u_{b}(\pi) | 
\mathcal{U}_{(\pi\rightarrow 0)} 
|\tilde{u}_{c}(0)\rangle \langle u_{c}(0) | 
\hat{C}_{2x} 
|\tilde{u}_{n}(2\pi)\rangle]
\\
\nonumber
&=& \sum_{a,b,c\in\textrm{occupied}} \det [\langle u_{m}(0) |  \mathcal{U}_{(0\rightarrow \pi)} |\tilde{u}_{a}(\pi)\rangle
\mathcal{B}^{-1}_{ab}(\pi) \langle u_{b}(\pi) | 
\mathcal{U}_{(\pi\rightarrow 0)}  
|\tilde{u}_{c}(0)\rangle \mathcal{B}_{cn}(0)]
\\
\nonumber
&=&
\det [ \mathcal{B}^{-1}(\pi)\mathcal{B}(0)] \times \det [\mathcal{U}_{(0\rightarrow \pi)} \mathcal{U}_{(\pi\rightarrow 0)} ]
\\
\nonumber
&=&
\det [ \mathcal{B}^{-1}(\pi)\mathcal{B}(0)] 
\\
\nonumber
&=&
\prod_{n\in\textrm{occupied}} \zeta_{n}(0)\zeta_{n}(\pi).
\eea
We have used the notation $|\tilde{u}(k)\rangle \equiv \Sigma_z |u(k)\rangle$ for the simplicity. Finally, we have expressed the Berry phase as a function of $\hat{C}_{2x}$ eigenvalues. Since $C_{2x}^2=1$ for the HP boson, the eigenvalues can only take a value of $\pm 1$. As a result, the Berry phase is quantized to either $0$ or $\pi$, which gives rise to the $\mathbb{Z}_2$ classification of the topological phase.

\end{widetext}
\end{document}